# Wavelength-dependent photo-creep in halide perovskite single crystals

Ruitian Chen[1], Jincong Pang[2], Lizhong Lang[1], Jiaze Wu[1], Mingyu Xie[1], Shuo Yang[3], Kaiqi Qiu[1], Tobin Filleter[3], Kai Huang[1], Guangda Niu[2], Jiang Tang[2], Yu Zou[1,*]

[1]Department of Materials Science and Engineering, University of Toronto, Toronto, ON M5S 3E4, Canada
[2]Wuhan National Laboratory for Optoelectronics and School of Optical and Electronic Information, Huazhong University of Science and Technology, Wuhan, 430074, China
[3]Department of Mechanical & Industrial Engineering, University of Toronto, Toronto, ON M5S 3G8, Canada

*Corresponding author. E-mail address: mse.zou@utoronto.ca (Yu Zou)

## Abstract

Halide perovskites are promising optoelectronic materials, but their time-dependent permanent deformation under illumination (i.e., photo-creep) is poorly understood, limiting their mechanical stability. Here we report wavelength-dependent photo-creep phenomena in $CsPbBr_3$ and $FAPbBr_3$ single crystals, studied by constant-load nanoindentation under controlled light with various wavelengths. Compared with creep in dark, continuous green light (near-bandgap) suppresses creep by 19% in $CsPbBr_3$ and 10% in $FAPbBr_3$, whereas violet (far above-bandgap) light enhances creep by 16% in $CsPbBr_3$ and 8% in $FAPbBr_3$. In contrast, when light is onset during creep, blue light enhances creep most prominently, whereas green light exhibits minimal influence. Such photo-creep behavior in halide perovskites are distinct with photo-plasticity phenomenon in conventional semiconductors. By combining the photoluminescence and photocurrent measurements, we unveil that ion migration promotes dislocation climb and creep, while carrier trapping suppresses dislocation glide and related creep in halide perovskites. Such competition between carrier trapping and ion migration tuned by wavelength governs the photo-creep response. Our findings uncover a photomechanical effect in halide perovskites and highlight how coupled carrier and ion dynamics under illumination affect their device reliability.

1. **Introduction**

Halide perovskites have emerged as highly efficient optoelectronic materials, including applications in solar cells, photodetectors, light-harvesting systems, and photocatalysts [1–7]. However, their long-term stability under operating conditions remains a critical challenge [8]. In particular, the lifetime of halide perovskite solar cells (PSCs) decreases significantly under illumination compared with that in darkness [9], and the PSCs show rapid efficiency loss under solar irradiation [10]. In halide perovskites, light-induced phase segregation and ion migration degrade the crystal structure, and mechanical failure often accompanies with defect formation [8,11]. A fundamental understanding of how illumination affects perovskite mechanical stability is therefore critical to reveal instability mechanisms and to design more durable optoelectronic devices.

Light is known to influence the mechanical properties of many semiconductor materials, a phenomenon often termed the photoplastic effect (PPE) [12–17], which describes the plastic deformation behavior under various light conditions [18–23]. Photo-nanoindentation has emerged as a powerful tool for studying PPE, enabling the detection of key deformation features such as hardness, pop-in events, and creep responses [24,25]. For example, nanoindentation of ZnS under 365 nm light showed ~17% higher hardness than in the dark, correlated with reduced dislocation mobility [26]. Photo-creep, referring to light-induced, time-dependent plastic deformation under constant load, has also been reported. Nanoindentation creep tests of ZnO demonstrated the suppressive effect of light on dislocation motion [27,28]. However, in halide perovskites ionic defects are mobile even at room temperature [29,30], so light may drive ion migration as well as electron excitation [31–34] and influence dislocation mobility [35]. Consequently, carrier and ion dynamics in halide perovskites become more complicated under illumination than those in conventional semiconductors. Despite halide perovskite single crystals have been studied using nanoindentation [36–38], the effect of illumination on their plasticity and creep behavior over a wide range of light wavelengths has not been explored.

In this work, we perform nanoindentation on single crystals of a typical all-inorganic halide perovskite $CsPbBr_3$ and a typical organic-inorganic halide perovskite (OIHP) $FAPbBr_3$ ($FA^+$ = $CH(NH_2)_2^+$) under varied illumination conditions. We quantify indentation creep, hardness, and pop-in events under different wavelengths to map out the wavelength-dependent mechanical response. In addition, we measure the photoluminescence (PL) spectroscopy and photocurrent of both crystals to probe the carrier and ion dynamics under illumination. By correlating these mechanical and optical measurements, we develop a mechanistic framework linking photo-induced electron–ion interactions with dislocation-mediated deformation in halide perovskites.

## 2. Results

### 2.1. Creep under continuous illumination

Figure 1 (a) illustrates the schematics of photo-nanoindentation testing on a $CsPbBr_3$ (bandgap of 2.3 eV) or $FAPbBr_3$ (bandgap of 2.2 eV) single crystal under the light with different wavelengths (Table 1 in Method, UV-Vis absorption spectra of materials and LED wavelengths in Figure S1, Supplemental materials Sec.1). No cracks are observed in the AFM topographic images of indents after creep tests for both crystals (Figure 1 (d)). Raman shifts and intensities of both the as-grown samples and the indented regions indicate that no phase transformation occurs after creep measurements (Figure S2, Supplemental materials Sec. 2).

We apply the constant load hold (CLH) method to measure the creep behavior of $CsPbBr_3$ and $FAPbBr_3$ single crystals using nanoindentation (Supplemental materials Sec. 3). The creep depth ($h_d$) as a function of holding time ($t_d$) is described as follows [39,40]:

$$h_d = \beta \cdot (t_d)^m \qquad (1)$$

where β and m are fitting parameters. The creep stress exponent (*n*) is defined as follows:

$$n = \frac{\partial(\ln \dot{\varepsilon})}{\partial(\ln H)} \qquad (2)$$

where $\dot{\varepsilon}$ is the indentation strain rate and *H* is the hardness.

For both crystals, the creep depth under violet light (far above-bandgap) is larger than that measured in darkness, whereas green light (near-bandgap) leads to a reduced creep depth (Figure 1 (b) and (c)). The creep depth under blue light (above-bandgap) lies between those observed under violet and green lights, while red light (below-bandgap) produces no measurable influence compared with darkness (Figure S3, Supplemental materials Sec. 3). These results indicate that illumination with photon energies near or above the perovskite bandgap exerts competing and wavelength-dependent effects on creep behavior. This trend is also reflected in the $n$ values shown in Figure 1 (e) and (f) (Representative fittings of the $n$ values in Figure S4, Supplemental materials Sec. 3). The statistical significance of the differences in $n$ values is verified using Welch's $t$-test to support the comparisons (Supplemental materials Sec. 7) [41].

To investigate the influence of light intensity on creep behavior, three different light intensities are applied for each wavelength during the creep tests. The variation trends of $n$ values shows that the change of $n$ values becomes saturated with increasing light intensity (Figure S5, Supplemental materials Sec. 4). To minimize the influence of light intensity on comparisons among different wavelengths, the results shown in Figure 1 are measured at relatively high light intensities where the $n$ values are stable. The creep depth of $FAPbBr_3$ is almost twice that of $CsPbBr_3$, indicating that $CsPbBr_3$ exhibits higher creep resistance than $FAPbBr_3$. The $n$ values of $CsPbBr_3$ are in the range of 8-13 under different illumination conditions, whereas the $n$ values of $FAPbBr_3$ are approximately from 3 to 5. These $n$ values of both crystals suggest that the dominant mechanism for creep deformation is associated with dislocation glide and dislocation climb [37,42].

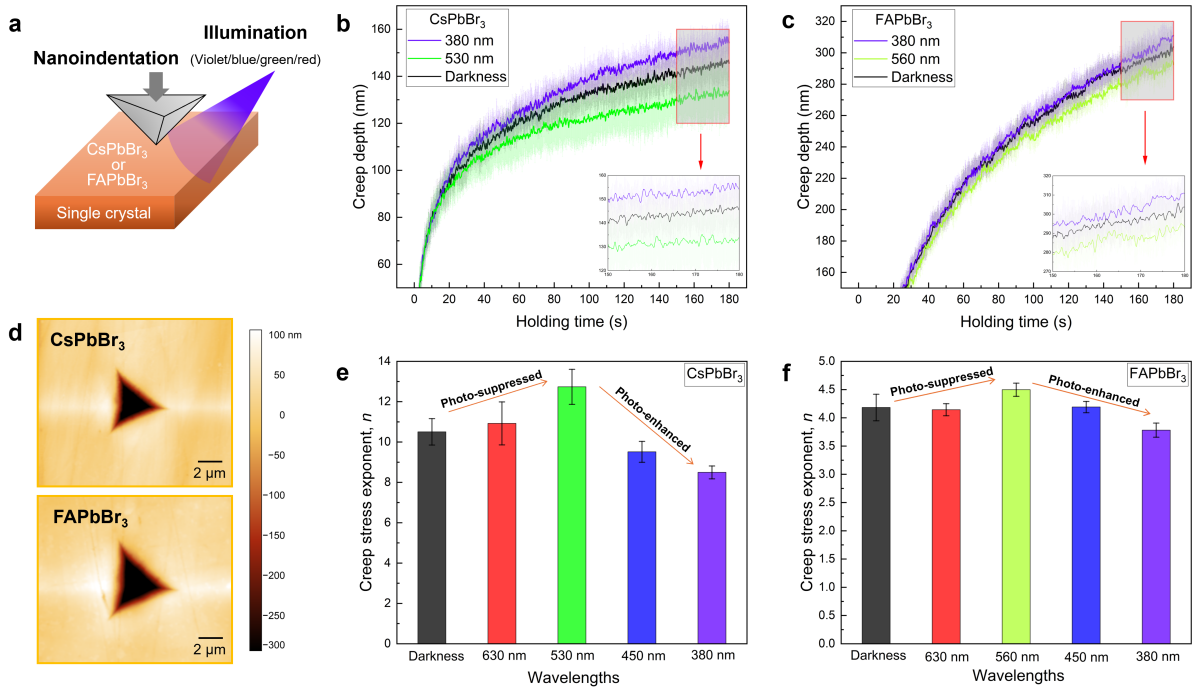

Figure 1. (a) Schematic illustration of photo-nanoindentation. Creep depth as a function of holding time of (b) CsPbBr$_3$ and (c) FAPbBr$_3$ single crystals under violet and green lights, compared with measurements in darkness. (d) AFM topographic images of indents after creep tests. Creep stress exponents ($n$) of (e) CsPbBr$_3$ and (f) FAPbBr$_3$ single crystals under different illumination conditions. (Light intensities: 380 nm, 18.0 mW cm$^{-2}$; 450 nm, 16.5 mW cm$^{-2}$; 530 nm, 13.4 mW cm$^{-2}$; 560 nm, 3.7 mW cm$^{-2}$; 630 nm, 17.2 mW cm$^{-2}$)

## 2.2. Illumination onset during creep

To further explore the influence of the effect of illumination during creep on these single crystals, creep tests are first conducted in the darkness, after which light is introduced during the load-holding stage, as illustrated in Figure 2 (a). The creep depth at the load-holding time of 30 s is taken as the reference value, and the light is switched on at 60 s. In addition, the creep depth change in darkness throughout the entire load-holding period is also shown for comparison.

Unlike the behavior observed under continuous illumination, blue light enhances creep more prominently than violet light when illumination is onset during creep, whereas green light has little effect on the creep change after it is turned on, as presented in Figure 2 (b-g). Besides, red light exhibits no influence since it is below the bandgap (Figure S6, Supplemental materials Sec. 5). During the prior loading and load-holding stage in darkness, dislocations have been nucleated and accumulated beneath the indenter before illumination. These distinct responses, observed between creep under continuous illumination and illumination onset during creep, indicate that variations in the initial dislocation density at the moment of illumination play an important role in the photo-creep behavior of halide perovskite single crystals.

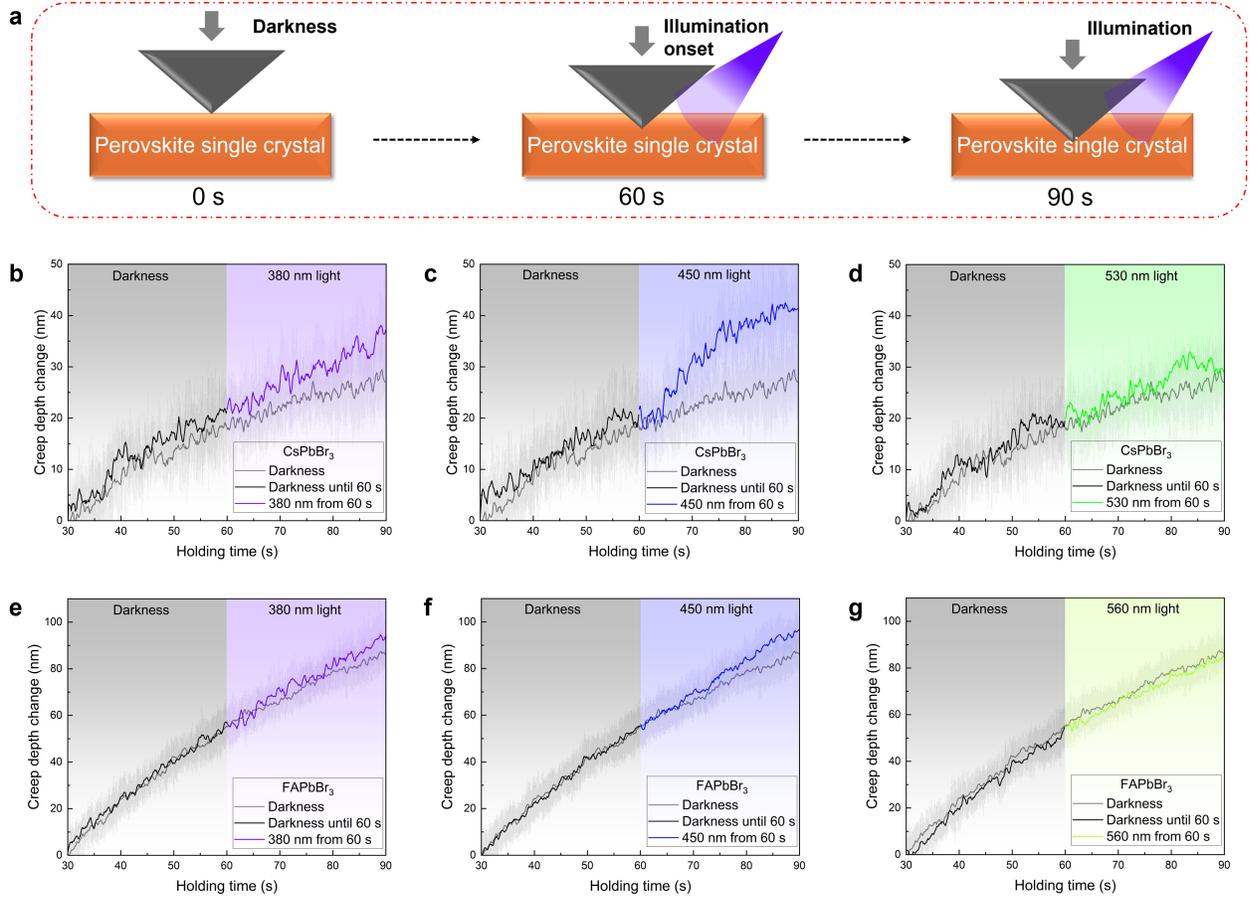

Figure 2. (a) Schematic of creep tests for illumination onset during creep. Representative creep depth changes over holding time of CsPbBr$_3$ and FAPbBr$_3$ single crystals under (b, e) violet light, (c, f) blue light, and (d, g) green light. (Light intensities: 380 nm, 18.0 mW cm$^{-2}$; 450 nm, 16.5 mW cm$^{-2}$; 530 nm, 13.4 mW cm$^{-2}$; 560 nm, 3.7 mW cm$^{-2}$)

### 2.3. Nanoindentation hardness and pop-in under illumination

To compare the photo-induced creep to plasticity in these single crystals, we also investigate the influence of illumination on their hardness and pop-in events. The hardness change ($\Delta H$) for each wavelength is defined as the following equation:

$$\Delta H = \frac{H_{\text{Light}} - H_{\text{Darkness}}}{H_{\text{Darkness}}} \times 100\% \qquad (3)$$

where $H_{\text{Light}}$ is the hardness tested under illumination and $H_{\text{Darkness}}$ is the hardness tested in darkness (Figure S7, Supplemental materials Sec. 6). In contrast to the creep behavior, only green light leads to an increase in hardness of both crystals (about 4% in $\Delta H$), while other wavelengths exhibit negligible effects compared with measurements in darkness (less than $\pm 1.5\%$ in $\Delta H$), as shown in Figure 3 (a) and (d). The statistical significance of the hardness variations is also verified using Welch's *t*-test (Supplemental materials Sec. 7). Representative load-depth curves in hardness tests also show the hardening effect of green lights (Figure S8, Supplemental materials Sec. 6). Light intensity has little significant influence on hardness (Figure S9, Supplemental materials Sec. 6), and the data presented in Figure 1 and Figure 3 are measured at the same light intensities correspondingly. The indentation strain rate of hardness tests is much higher than that in the creep tests, so ion migration does not play significant roles [43]. Consequently, hardness primarily reflects the resistance to dislocation glide and plastic deformation [26,38], and only the light photon energy near the bandgap has the suppression effect on dislocation glide in halide perovskite single crystals.

Pop-in events, which are identified as sudden displacement bursts appearing as plateaus in the loading curves, are generally associated with crack formation, phase transformation, or dislocation nucleation [28,38,44,45]. The load-depth curves of $CsPbBr_3$ and $FAPbBr_3$ single crystals exhibit multiple pop-in events within the first 100 nm of nanoindentation depth, as shown in Figure 3 (b) and (e). In the halide perovskite single crystals, the first pop-in event is regarded as the onset of plastic deformation (i.e. dislocation nucleation), because neither cracking nor phase transformation is observed. The cumulative probability distributions as a function of the maximum shear stress at the first pop-in event (Supplemental materials Sec. 8) reveal that no clear trend or systematic variation is observed between measurements conducted under different illumination conditions and in darkness, as shown in Figure 3 (c) and (f), indicating that dislocation nucleation is not significantly influenced by illumination.

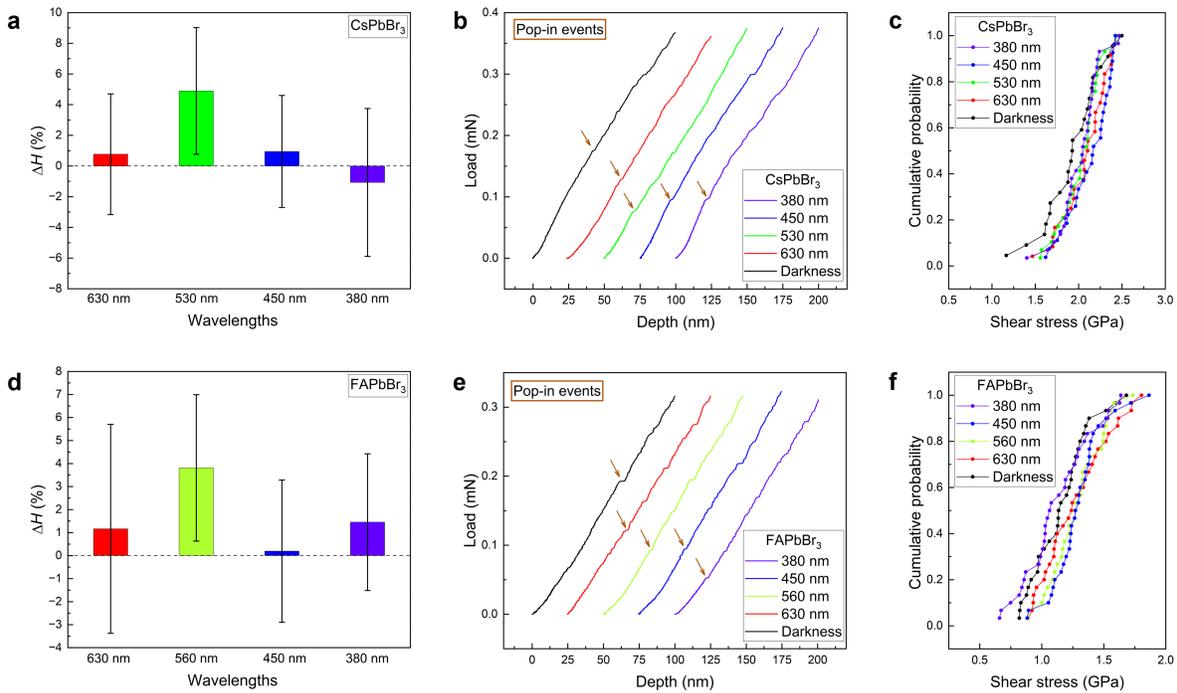

Figure 3. The hardness change ($\Delta H$), pop-in events in load-depth curves, and cumulative probability as a function of maximum shear stress at the first pop-in event of (a-c) CsPbBr$_3$ and (d-f) FAPbBr$_3$ single crystals under different illumination conditions. Curves are offset for clarity, and some pop-in events are highlighted by arrows in (b, e). (Light intensities: 380 nm, 18.0 mW cm$^{-2}$; 450 nm, 16.5 mW cm$^{-2}$; 530 nm, 13.4 mW cm$^{-2}$; 560 nm, 3.7 mW cm$^{-2}$; 630 nm, 17.2 mW cm$^{-2}$)

## 2.4. Carrier and ion dynamics under illumination

To understand the carrier and ion dynamics under illumination, PL spectra of $CsPbBr_3$ and $FAPbBr_3$ single crystals are measured under excitation with violet, blue, and green lights, as shown in Figure 4 (a) and (b). Under green light excitation, the PL peaks exhibit a pronounced red shift relative to the bandgap in both crystals. In contrast, under violet and blue light excitation, the PL peaks are located close to the bandgap, with only a slight red shift observed under violet light. Moreover, the full width at half maximum (FWHM) increases sequentially from green to blue and to violet light excitation. These observations indicate distinct carrier and ion dynamics in these single crystals under different illumination conditions.

When the light photon energy reaches or exceeds the bandgap energy, electrons are excited from the valence band (VB) to the conduction band (CB). Meanwhile, photo-induced ion migration is activated in halide perovskites [32]. After excitation, electrons may be trapped at defect sites prior to radiative recombination [46]. For different excitation energies, the competition between electron trapping and ion migration leads to distinct recombination pathways, resulting in variations in PL peak positions and FWHM, as illustrated in Figure 4 (c-e). Under green light excitation, electrons are preferentially trapped by dislocations [31], giving rise to deep-trap emission and a pronounced red shift of the PL peak. In contrast, under blue light excitation, electrons can be promoted to higher energy states in the conduction band, and ion migration becomes more pronounced. This enhanced ion migration facilitates the partial healing of dislocation trap sites, leading predominantly to near-band-edge emission [31]. Violet light possesses higher photon energy and a shallower absorption depth, which can induce increased surface defect generation and accelerated ion migration [33,46]. Consequently, excited electrons are more likely to be trapped by surface defect states, resulting mainly in shallow-trap emission. In addition, as ion migration intensifies and surface defect density increases under illumination with higher light photon energy, the recombination processes become more complex, leading to broader PL spectra and larger FWHM values.

Photocurrent results shown in Figure 4 (f) and (g) indicate that green light induces a higher current density per flux in both crystals, which is consistent with its greater absorption depth (Supplemental materials Sec. 9). $FAPbBr_3$ exhibits a current density approximately one order of magnitude higher than that of $CsPbBr_3$, which can be attributed to the suppressed nonradiative recombination in $FAPbBr_3$, as evidenced by its higher PL intensity (Figure S10, Supplemental materials Sec. 10), Moreover, $FAPbBr_3$ exhibits pronounced photocurrent hysteresis under illumination, which arises from its enhanced ion migration. The hydrogen bonding between the organic cation and the inorganic framework can be modified by light, increasing the mobility of $FA^+$ and thereby amplifying ion-driven effects [47–49].

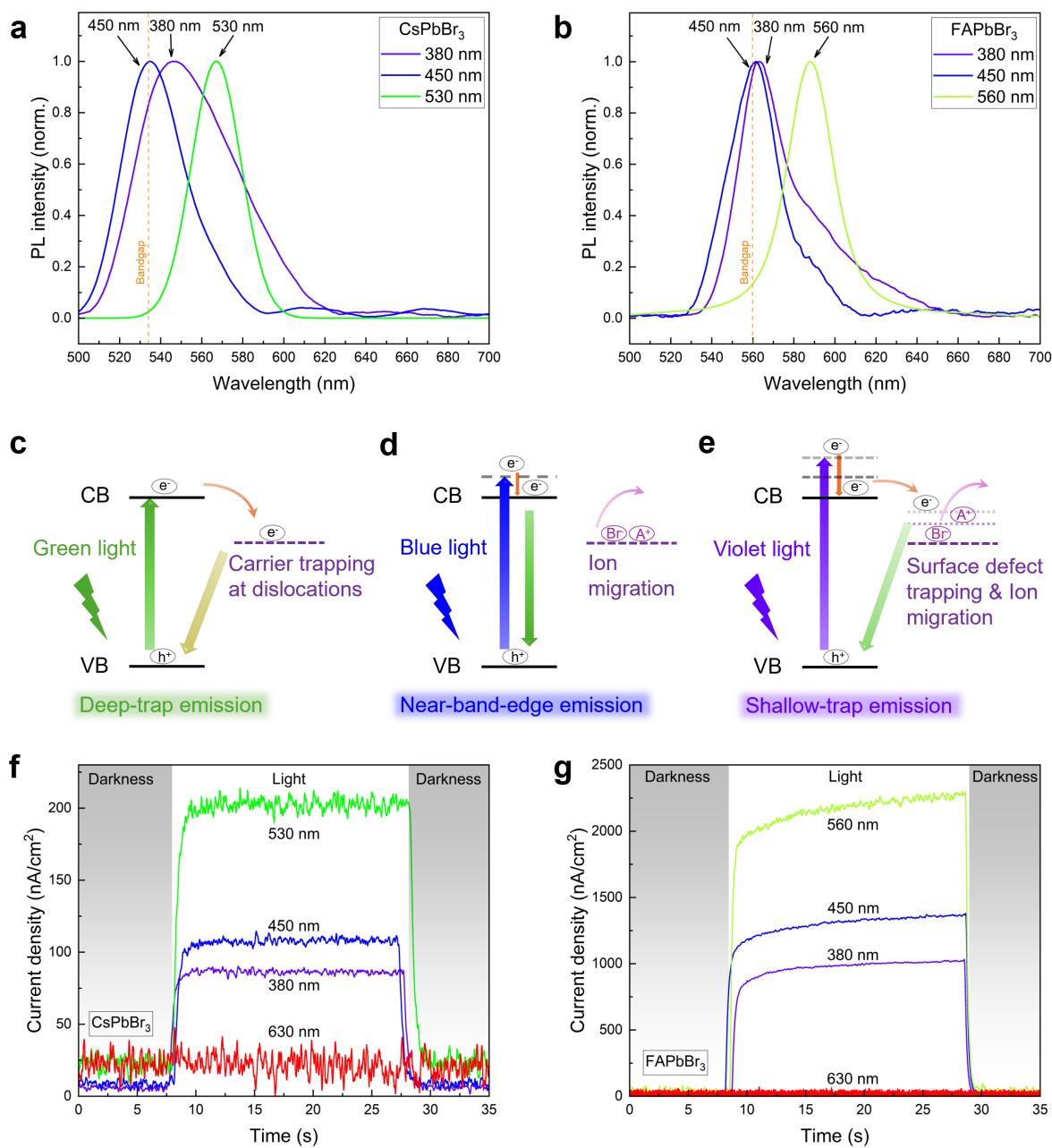

Figure 4. Normalized PL spectra of (a) CsPbBr$_3$ and (b) FAPbBr$_3$ single crystals under excitation of different lights. (c-e) Schematic illustrations of photo-induced carrier and ion interactions with dislocations (VB: valence band; CB: conduction band; e$^-$: electron; h$^+$: hole; A$^+$: Cs$^+$ or FA$^+$). Photocurrent results of (f) CsPbBr$_3$ and (g) FAPbBr$_3$ single crystals under different illumination conditions, with current densities normalized to a photon flux of 10$^{15}$ photons s$^{-1}$ cm$^{-2}$.

## 3. Discussion

### 3.1. Distinct creep mechanisms under different illumination conditions

During the loading stage of nanoindentation creep, dislocations nucleate rapidly as the indentation depth increases. During the load-holding stage, subsequent dislocation climb and dislocation glide dominate the creep behavior. Dislocation glide is primarily controlled by plasticity, while dislocation climb is governed by ion migration [50]. Figure 5 schematically illustrates the wavelength-dependent dislocation dynamics both under continuous illumination and for illumination onset during creep. Under continuous illumination (Figure 5 (a)), dislocation nucleation is not significantly affected by light, as evidenced by the similar characteristics of pop-in events (Figure 3). In contrast, dislocation climb is progressively enhanced with increasing light photon energy, because higher energy accelerates ion migration as reflected by the broader PL spectra. Ion migration facilitates dislocation climb by promoting defect accumulation and fast diffusion along dislocation cores [42,50]. Meanwhile, pronounced carrier trapping under green light suppresses dislocation glide, as evidenced by the red shift of PL peaks. Such carrier–dislocation interactions can stabilize bond reconstructions and enhance stress fields at dislocation cores, and these reconstructed bonds should be broken during dislocation glide, thereby significantly restricting dislocation motion [23,26,51]. The competition between dislocation climb (dominated by ion migration) and dislocation glide (dominated by carrier trapping) ultimately governs the creep responses (Figure 5 (b)): (i) Under violet light, ion migration dominates, leading to enhanced creep deformation; (ii) Under green light, carrier trapping prevails, resulting in suppressed creep; (iii) Under blue light, these two effects are comparable, and the creep behavior lies between the two extremes.

The dislocation behavior for illumination onset during creep differs from that under continuous illumination (Figure 5 (c)). When light is switched on after a period of load holding, a higher density of dislocations has already nucleated, providing abundant diffusion pathways for ion migration [32,47]. As a result, the dominate factor governing ion migration at that moment is no longer the light photon energy, but rather the light absorption depth (Supplemental materials Sec. 9). Consequently, green light, which penetrates more deeply than blue and violet lights, promotes ion migration more effectively and thus enhances dislocation climb to a greater extent. This trend is opposite to the circumstance under continuous illumination. Meanwhile, dislocation glide remains influenced by carrier trapping under green light. The interplay between dislocation climb and dislocation glide leads to the strongest creep enhancement when blue light is introduced (Figure 5 (d)).

In halide perovskite single crystals, the influence of light on ion migration is modulated by the dislocation density. When dislocation density is relatively low, the scarcity of natural migration pathways renders ion motion primarily dependent on the light photon energy. Illumination with higher energy can effectively lower ion migration barriers, leading to a pronounced enhancement of ionic mobility. In contrast, when dislocation density is relatively high, the dense network of dislocations provides enough ion migration pathways [32,47], such that the overall migration is dominated by the volume of the crystal accessed by the light. In this case, variations in light photon energy play a secondary role, as ions can readily migrate along the dislocation network once activated.

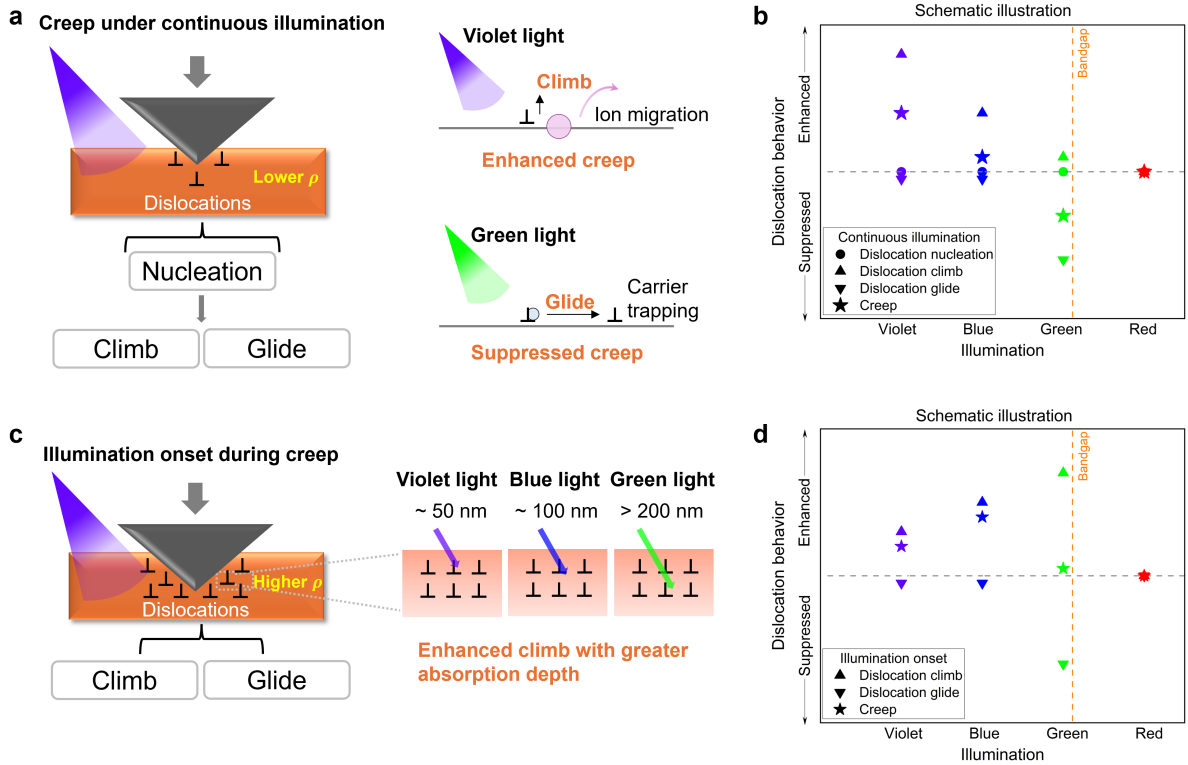

Figure 5. Schematics of the wavelength-dependent dislocation behavior in nanoindentation creep. (a, b) Under continuous illumination, violet light strongly promotes ion migration and enhances dislocation climb, whereas green light mainly induces carrier trapping and suppresses dislocation glide. (c, d) For illumination onset during creep, green light enhances dislocation climb more effectively due to its greater absorption depth.

## 3.2. Influence of A-site cations on photo-creep behavior

$FAPbBr_3$ exhibits a smaller creep stress exponent than $CsPbBr_3$ under identical conditions, indicating a more pronounced contribution of dislocation climb to the creep of $FAPbBr_3$ [38]. The larger size and lower symmetry of the $FA^+$ organic cation induce structural distortions in the perovskite framework [52]. Moreover, the activation energies for vacancy-assisted ion migration are reported to be 0.33 eV for $Br^-$ and 0.39 eV for $FA^+$ in $FAPbBr_3$, compared with 0.51 eV for $Br^-$ and 0.80 eV for $Cs^+$ in $CsPbBr_3$ [49,53,54]. These lower energy barriers of ion migration facilitate ion migration in $FAPbBr_3$, leading to enhanced dislocation climb and creep deformation.

The higher photocurrent density and suppressed nonradiative recombination in $FAPbBr_3$ indicate reduced defect formation and lower trap densities. As a result, variations in carrier and ion dynamics under different illumination conditions are less pronounced, leading to a more subtle creep response in $FAPbBr_3$ (less than 10% variation in creep exponents) compared with $CsPbBr_3$ (more than 15% variation in creep exponents) under continuous illumination. In addition, $FAPbBr_3$ exhibits a longer delay in creep depth change than $CsPbBr_3$ for illumination onset during creep, which correlates with its pronounced photocurrent hysteresis.

### 3.3. Comparisons between halide perovskites and conventional semiconductors

Table S7 (Supplemental materials Sec.11) summarizes the photo-induced creep and plasticity behavior in various semiconductors, including conventional ionic and covalent semiconductors from representative previous studies, as well as halide perovskites investigated in our work. For conventional semiconductors, studies of photomechanical effects have primarily focused on illumination near the bandgap. Notably, photoplastic phenomena differ substantially between ionic II-VI compounds and covalent III-V compounds [17]. In ZnS and ZnO, near-bandgap illumination enhances flow stress, hardness, and creep resistance [22,23,25–28], which is attributed to elevated Peierls barrier and reduced dislocation glide mobility as revealed by simulations [21,26,51,55,56]. However, there is a crossover from photo-induced hardening to softening in CdTe depending on the light intensity [57,58]. By contrast, near-bandgap illumination softens GaAs and GaP crystals and promotes dislocation mobility [59–62]. In our work, halide perovskites, with high ionicity,

similarly exhibit increased hardness and suppressed creep behavior under illumination near the bandgap. In addition, the maximum shear stress at the first pop-in event in halide perovskites is insensitive between illuminated and dark conditions, consistent with previous observations in ZnO [63].

For conventional semiconductors, illumination with light energies far above the bandgap generally exerts only subtle influences on their plasticity and creep behavior [25,26]. Similarly, in halide perovskites, hardness is primarily affected by illumination near the bandgap. However, their creep behavior responds not only to near-bandgap illumination but also to above-bandgap light, with competing effects depending on the wavelength. This wavelength-dependent creep response originates from the pronounced light-driven ion migration in halide perovskites, a process that is largely negligible in conventional semiconductors. Regarding the influence of light intensity, the variation in creep resistance and hardness in halide perovskites initially increases with increasing light intensity and subsequently reaches a rapid saturation. This trend is consistent with observations reported for ZnS [21,26].

Beyond single crystals, polycrystalline halide perovskite thin films are widely employed in solar cells and flexible electronics and also exhibit pronounced response to light illumination [7,8,34,47]. In conventional polycrystalline semiconductors, grain boundaries and surface defect states play a dominant role in governing mechanical behavior [17]. Consequently, investigating the photomechanical effects in polycrystalline halide perovskites is also of significant interest and the underlying mechanisms are expected to be more complicated. Overall, photo-creep provides opportunities for light-assisted processing and defect engineering at room temperature, while also playing a critical role in governing the long-term mechanical and operational stability of perovskite devices. These capabilities highlight the broader potential of leveraging photo-induced defect dynamics for multifunctional material design.

**4. Conclusions**

In summary, we have studied wavelength-dependent photo-creep in $CsPbBr_3$ and $FAPbBr_3$ single crystals using nanoindentation with *in situ* illumination. We find that the creep response is strongly tuned by light photon energy: Under continuous illumination, green (near-bandgap) light suppresses creep, whereas violet (far above-bandgap) light greatly enhances creep. When illumination is onset during creep, blue light enhances creep more prominently than violet light, whereas green light exhibits minimal influence. Hardness and pop-in measurements reveal that only green light reduces dislocation glide mobility, while dislocation nucleation is insensitive to illumination conditions. PL and photocurrent results elucidate the carrier and ion dynamics in halide perovskite single crystals, showing the behavior of carrier trapping and ion migration under illumination. We propose that photo-generated mobile ions promote creep by enabling dislocation climb, whereas trapped carriers hinder dislocation glide. The competition between these processes, modulated by wavelength, governs the overall photo-creep behavior. We have also analyzed the influence of A-site cations on photo-creep behavior in halide perovskites and compare our findings with conventional semiconductors. These findings improve our understanding of photo-induced mechanical degradation and have implications for the reliability of perovskite devices under operating conditions.

## 5. Methods

### 5.1. Synthesis of halide perovskite single crystals

CsPbBr$_3$ single crystals were synthesized from stoichiometric CsBr and PbBr$_2$ via the Bridgman method. FAPbBr$_3$ single crystals were grown from a precursor solution prepared by dissolving FABr and PbBr$_2$ in γ-butyrolactone. Experiments were conducted on the (101) plane of CsPbBr$_3$ single crystals and (100) plane of FAPbBr$_3$ single crystals. Detailed parameters for the synthesis of CsPbBr$_3$ and FAPbBr$_3$ single crystals and X-ray diffraction (XRD) results are as reported in our previous studies [11,38].

### 5.2. Nanoindentation

Nanoindentation tests were performed in the KLA iMicro Nanoindentation system equipped with a Berkovich diamond indenter tip (Synton-MDP) at room temperature [11,38]. The tip shape function was calibrated by testing the fused silica standard. The drift rates were controlled below 0.1 nm/s, and the sampling frequency was 110 Hz. Creep was measured by the CLH method that the indenter tip reached a target load, followed by holding the load for a certain period [39]. In our tests, the target load was 3.6 mN, and the holding time was 180 s. Hardness was measured by the Oliver and Pharr method in the continuous stiffness measurement mode with an indentation depth of 600 nm [64,65]. Pop-in events were collected from load-depth curves with an indentation depth of 100 nm. Multiple tests were repeated to ensure reliable results. Figure S11 (Supplemental materials Sec.12) illustrates the load-time schemes used for the nanoindentation creep and hardness tests.

### 5.3. LED lighting

Commercial LEDs with a specific light emission range were used as illumination source. The wavelengths of LEDs were calibrated using a Horiba spectrofluorometer (FluoroMax Plus), and light intensities were measured with a Si-based photodiode power sensor (S120VC, Thorlabs). Light intensities during indentation were controlled by the current of the power supply. Table 1 summarizes the light wavelengths used in photo-indentation testing.

Table 1. Light used in photo-nanoindentation testing.

| Light wavelength (energy) | Details |
|:---:|:---:|
| 380 nm (3.26 eV) | Violet (far above-bandgap) |
| 450 nm (2.76 eV) | Blue (above-bandgap) |
| 530 nm (2.34 eV) | Green (near-bandgap) |
| 560 nm (2.21 eV) | Green (near-bandgap) |
| 630 nm (1.97 eV) | Red (below-bandgap) |

### 5.4. Atomic force microscopy, Raman microspectroscopy, and UV-Vis absorption spectroscopy

Surface topology of the indents was measured using an Asylum Cypher S atomic force microscope (Oxford Instruments) in a tapping mode with a Ti-Ir-coated silicon tip (ASYELEC.02-R2). Raman spectra were measured using a Raman microscope spectrometer (Bruker SENTERRA) at a laser wavelength of 785 nm and intensity of 1 mW. The measurements were carried out at room temperature with 10 s integration time. UV-Vis absorption spectra of halide perovskite single crystals were measured using a Lambda 365 Spectrometer (PerkinElmer) between 350 nm and 750 nm. The measurements were carried out at room temperature, with a 1 nm interval, 0.1 s integration time, and 2.0 nm bandwidth.

### 5.5. Photoluminescence spectroscopy and photocurrent measurement

PL spectra of halide perovskite single crystals were measured using a portable spectrometer (SENSE, Sarspec). The measurements were conducted at room temperature with a 0.5 nm interval and 0.1 s integration time. LEDs were used for excitation, and an optical filter (FGL530M/FGL550M, Thorlabs) was placed in front of the detector to filter out excitation light (Figure S12 (a), Supplemental materials Sec.12). The photocurrent measurements were performed using a P118 nA level current analyzer (Smart power). Two silver electrodes were brushed on both sides of halide perovskite single crystals (Figure S12 (b), Supplemental materials Sec.12). The voltage applied on each sample was set to a constant value.

**Acknowledgements**

R.C. and Y.Z. acknowledge the financial support from the Discovery Grants Program of the Natural Sciences and Engineering Research Council of Canada (NSERC) RGPIN-2018-05731, the Ontario Early Researcher Award, and the Canada Foundation for Innovation (CFI) - Evans Leaders Fund (JELF) Number 38044. The authors thank Mingqiang Li for the assistance of experiments.

**Authorship contributions**

**Ruitian Chen:** Conceptualization, Methodology, Investigation, Data curation, Writing – original draft, Writing – review & editing. **Jincong Pang:** Methodology, Resources, Writing – review & editing. **Lizhong Lang:** Methodology, Data curation, Writing – review & editing. **Jiaze Wu:** Investigation, Data curation, Writing – review & editing. **Mingyu Xie:** Investigation, Data curation, Writing – review & editing. **Shuo Yang:** Investigation, Writing – review & editing. **Kaiqi Qiu:** Investigation, Writing – review & editing. **Tobin Filleter:** Resources, Writing – review & editing. **Kai Huang:** Resources, Writing – review & editing. **Guangda Niu:** Resources, Writing – review & editing. **Jiang Tang:** Resources, Writing – review & editing. **Yu Zou:** Funding acquisition, Supervision, Resources, Writing – review & editing.

### Declaration of interests

The authors declare no competing interests.

### Data availability

Data will be made available upon request.

### Supplemental materials

Supplemental materials are provided in a separate file.